\documentclass[english,aps,prb,showpacs,floats,twocolumn]{revtex4}
\usepackage{amssymb}
\usepackage[latin9]{inputenc}
\usepackage{babel}
\usepackage{graphicx}
\usepackage{hyperref}
\usepackage{amsmath}
\input{tcilatex}
\begin{document}
\title{Optimal switching of a nanomagnet assisted by microwaves}
\author{N. Barros$^{1}$, M. Rassam$^{1}$, H. Jirari$^{2}$, and H. Kachkachi$^{1}$}
\address{
$^{1}$LAMPS, Universit de Perpignan Via Domitia, 52 Avenue Paul Alduy, 66860 Perpignan Cedex,
France.\\
$^{2}$LPMMC, C.N.R.S-UMR 5493, Universit\'e Joseph Fourier, BP 166, 38042 Grenoble-Cedex 9, France}
\begin{abstract}
We develop an efficient and general method for optimizing the microwave field that achieves magnetization
switching with a smaller static field.
This method is based on optimal control and renders an exact solution
for the $3D$ microwave field that triggers the switching of a nanomagnet with a given anisotropy
and in an oblique static field.
Applying this technique to the particular case of uniaxial anisotropy, we show that the optimal
microwave field, that achieves switching with minimal absorbed energy, is modulated both in frequency and in magnitude. Its
role is to drive the magnetization from the metastable equilibrium position towards the saddle point and then damping induces the
relaxation to the stable equilibrium position.
For the pumping to be efficient, the microwave field frequency must match at the early stage of
the switching process the proper precession frequency of the magnetization, which depends on the
magnitude and direction of the static field. 

We investigate the effect of the static field (in amplitude and direction) and of damping
on the characteristics of the microwave field. We have computed the switching curves in
the presence of the optimal microwave field. The results are in qualitative agreement with $\mu$-SQUID experiments on
isolated nanoclusters.
The strong dependence of the microwave field and that of the
switching curve on the damping parameter may be useful in probing damping in various nanoclusters.
\end{abstract}


\maketitle

\section{Introduction}

Fine magnetic clusters offer tremendous challenges
both in the area of fundamental science and practical applications.
The main reasons for this impetus are the novel features related with their small size, such as the
possibility of high density storage, short-time switching \cite{heetal10prb} and fast read-write
processes. On the other hand, the small size is a drawback in many regards. The
energy barrier in these systems is too small to ensure a reasonable stability, in a given energy
minimum, that is necessary for practical applications at room temperature, \textit{e.g.}, magnetic recording.
This is the problem of superparamagnetism.
A possible way out would be to use materials with high anisotropy and thus ensuring a high energy
barrier. A consequence of this is that high values of the writing (or switching) fields are
required. However, it is still unclear how to devise such high fields operating on the scale of
nanoclusters while avoiding the ensuing noise.
In order to keep the size small, the energy barrier high, and the switching field small,
other routes are explored and a promising one among them is provided by microwaves.
Microwave-assisted magnetization switching in various magnetic systems, such as thin films, has
been investigated by many groups \cite{OtherSystems}.
In fact, we have at hand a more general and fundamental issue, namely the problem of getting a
system out of an energy minimum by nonlinear resonance. This has previously been addressed in many
areas of physics and chemistry, especially in the context of atomic physics. 
For example, Liu et al. \cite{liuetal95prl} have studied the dissociation of diatomic molecules by a
chirped infrared laser pulse and showed that this process requires a much lower threshold laser
intensity to achieve dissociation.
The quantum regime \cite{yualiu98pra} has been studied in terms of energy-ladder climbing and gives
very similar results. Experimental evidence of this process has been provided by the dissociation of
HF molecules using a sub-nanosecond frequency modulated laser pulse
\cite{marcusetal}.

According to the classical theory of auto-resonance or the quantum theory of
ladder-climbing \cite{chelkowskietal90prl, schirmeretal01prb, marcusetal}, exciting an oscillatory
nonlinear system to high energies is possible by a weak \textit{chirped frequency} excitation.
Moreover, trapping into resonance followed by a (continuing and stable) \textit{phase-locking} with
the drive is possible if the driving frequency chirp rate is small enough. It has also been shown
that a slow passage through and capture into resonance yields efficient control of 
the energy of the driven system.
Incidentally, an important theoretical result is that the precise form of the time dependence of
the oscillating field is not essential for the process to succeed.

For nanoclusters, it has been shown in previous works how a monochromatic microwave (MW) pulse can, by means of
a non-linear resonance, substantially reduce the required static field needed to reverse the
magnetization of an individual nanoparticle.  
Indeed, it has been demonstrated using the $\mu$-SQUID technique on a 20-nm cobalt particle
\cite{microSQUID} that adding an MW field with given rising time, duration, and
frequency, on top of the static (DC) magnetic field, the switching of the nanocluster is possible at
a static field lower than the Stoner-Wohlfarth (SW) switching field and within a time interval of the order of
a nanosecond. 
The switching curves, or the so-called SW astroids, obtained in these measurements present some
irregular features implying that the reduction of the switching field is not uniform, as is the
case with thermal effects. The global features depend on several physical parameters, such as the
MW field pulse duration, its rising time, and its frequency, the DC field amplitude, and the damping
parameter \cite{microSQUID}.
The SW astroid obtained with its peculiar features seems to bear the
fingerprints of the nanocluster and its underlying characteristics such as its potential energy.
Moreover, the strong dependence of these features on the damping parameter might be used to estimate
the latter in such clusters.

On the theory side, several works have been devoted to the understanding of the magnetization
dynamics, and in particular its reversal, under the effect of a time-dependent magnetic field. 
The theoretical work may be divided into two kinds. The first deals with the effect of a given
MW field with a given polarization \cite{denisovetal06prl, bertottietal02jap} while the second
seeks optimal strategies for achieving the magnetization switching \cite{sunwang06prb}. In
particular, a few works, \textit{e.g.} \cite{mayergoyzetal}, assume a given dynamics for
the magnetization and attempt to determine the MW field that realizes it.

In the present work, we use a general method borrowed from the optimal control theory
\cite{bryho75tf, pierceetal88pra, jirpot} and apply it to the switching of a
nanocluster. This method renders an exact solution for the MW field vector necessary for the
switching of a nanomagnet with a given potential energy (comprising anisotropy and an oblique static
field).
The standard formulation of this method consists in minimizing a cost functional using the
conjugate gradient technique. The latter is known to be a local-convergence method and thus
renders a solution that is rather sensitive to the initial guess.
In order to acquire global convergence and thus have solutions for the MW field that are stable
with respect to a change in the initial conditions, we have supplemented the conjugate gradient
routine by a global search using the Metropolis algorithm and simulated annealing.
Then, we have applied our algorithm to a nanomagnet in the macrospin approximation with
uniaxial anisotropy and oblique DC magnetic field. We have investigated the effect of the latter (both in
direction and magnitude) and of damping on the characteristics of the MW field.
Then, we computed the limit-of-metastability (or switching) curves for different (small) values of damping. 

\section{Method of optimal control applied to nanomagnets}
One of our objectives here is to develop a general method that allows
us to solve the following inverse problem: what is the optimal time-dependent
magnetic field under which the magnetization of a nanoparticle, with
given potential energy, switches from a given initial state to a given
final prescribed target state? After formulating this method we apply
it to the case of a macrospin, or a nanoparticle in the SW
approximation, in an energy potential composed of uniaxial anisotropy
and a Zeeman contribution from an oblique DC field.

The method we propose is borrowed from the optimal control theory.
The main idea is to start with an arbitrary MW magnetic field $\mathbf{h} _{\mathrm{AC}}(t)$
with its three components $h_{\mathrm{AC}} ^{\alpha},\alpha=x,y,z$, that we call
the \textit{control field}, in addition to the static magnetic field
and anisotropy field. We then determine $\mathbf{h} _{\mathrm{AC}}(t)$ that triggers
the switching of the cluster's magnetization between two given states
within a prescribed interval of time.

\subsection{Model}
Consider a nanomagnet in the macrospin approximation where
its magnetic state is represented by a macroscopic magnetic moment
$\mathbf{m}=\mu_{s}\,\mathbf{s},$ where $\mu_{s}$ is its magnitude
and $\mathbf{s}$ its direction with $\left\vert \mathbf{s}\right\vert =1$.
In this approximation, the relevant terms entering the energy $E$
of the nanomagnet are the magneto-crystalline anisotropy and the Zeeman
energy. The applied DC (static) field $\mathbf{H} _{\mathrm{DC}}$ is assumed to
point in an arbitrary direction $\mathbf{e}_{h}=\mathbf{H} _{\mathrm{DC}}/H _{\mathrm{DC}}$. Using the 
convention $\mu _0 = 1$ so that the magnetic fields are expressed in Tesla, 
one then defines the effective field
\begin{equation}
\mathbf{H}_{\mathrm{eff}}=-\frac{1}{\mu_{s}}\frac{\delta E}{\delta\mathbf{s}}\label{eq:EffField}
\end{equation}
and writes the damped Landau-Lifshitz equation in the Gilbert form (LLE) that governs
the dynamics of $\mathbf{s}$, assuming that the module $\mu_{s}$
remains constant,
\begin{equation}
\frac{1}{\gamma}\frac{d\mathbf{s}}{dt}=-\mathbf{s}\times\mathbf{H}_{\mathrm{eff}}-\alpha\,\mathbf{s}
\times\left(\mathbf{s}\times\mathbf{H}_{\mathrm{eff}}\right)\label{eq:OSPDampedLLE}
\end{equation}
where $\gamma \simeq1.76\times10^{11}$ (T.s)$^{-1}$ is the gyromagnetic
factor and $\alpha$ the phenomenological damping parameter (taken here in the weak
regime).

We measure all applied fields in terms of the anisotropy field
\begin{equation}
H_{a}=\frac{2KV}{\mu _s}\label{eq:AnisotropyField}
\end{equation}
 and in particular we define the reduced effective field
\begin{equation}
\mathbf{h}_{\mathrm{eff}}\equiv\frac{1}{H_{a}}\mathbf{H}_{\mathrm{eff}}=-\frac{
\delta\mathcal{E}}{\delta\mathbf{s}},\qquad\label{eq:EffRedField}
\end{equation}
with $\mathcal{E}\equiv E/\left(2KV\right)$. In terms of $\mathbf{h}_{\mathrm{eff}}$
LLE becomes
\begin{equation}
\frac{d\mathbf{s}}{d\tau}=-\mathbf{s}\times\mathbf{h}_{\mathrm{eff}}-\alpha\,\mathbf{s}
\times\left(\mathbf{s}\times\mathbf{h}_{\mathrm{eff}}\right)\label{eq:OSPDampedLLERed}
\end{equation}
where $\tau\equiv t/t_{s}$ is the dimensionless time and $t_{s}=1/(\gamma H_{a})$
the characteristic scaling time of the system. For instance, for a
cobalt particle of $3$ nm diameter \cite{jametetal01prl} with $K\simeq2.2\times10^{5}$
J.m$^{-3}$, $\mu_{s}\simeq3.8\times10^{-20}$ A.m$^{2}$ we have $H_{a}\simeq0.3$
T and $t_{s}\simeq1.9\times10^{-11}$ s.

\subsection{\label{subs:FOCP}Formulation of the optimal control problem}

\paragraph{General procedure}

The idea here is to introduce a control field $\mathbf{h} _{\mathrm{AC}}(\tau)\equiv \mathbf{H} _{\mathrm{AC}}/H_a$
and then seek its optimal form that allows for driving the magnetic
moment direction $\mathbf{s}$ from the given initial state $\mathbf{s}^{(i)}$
at time $\tau_{i}=0$ into the desired final state $\mathbf{s}^{(f)}$
at the given observation time $\tau_{f}$. Accordingly, we replace
in the LLE (\ref{eq:OSPDampedLLERed}) the (deterministic) field $\mathbf{h}_{\mathrm{eff}}$
by the total (time-dependent) field
\begin{equation}
\mathbf{\zeta}(\tau)=\mathbf{h}_{\mathrm{eff}}+\mathbf{h} _{\mathrm{AC}}(\tau).\label{eq:TotalEffectiveField}
\end{equation}

This results in the following equation of motion, which will be henceforth referred
to as the driven LLE (DLLE)
\begin{equation}
\mathbf{\dot{s}}=-\mathbf{s}\times\mathbf{\zeta(}\tau\mathbf{)}-\alpha\,\mathbf{s}\times\left(\mathbf{s
}\times\mathbf{\zeta(}\tau\mathbf{)}\right).\label{eq:DrivenLLE}
\end{equation}
 The field $\mathbf{h} _{\mathrm{AC}}(\tau)$ is then determined through the minimization
of a cost functional which, in the present case, may be written as
\begin{equation}
\mathcal{F}\left[\mathbf{s}(\tau),\mathbf{h} _{\mathrm{AC}}(\tau)\right]=\frac{1}{2}\left\Vert
\mathbf{s}(\tau_{f})\mathbf{-s}^{(f)}\right\Vert ^{2}+\frac{\eta}{2}\int\limits
_{0}^{\tau_{f}}d\tau\,\mathbf{h} _{\mathrm{AC}}^{2}(\tau)\label{eq:CostFunctional}
\end{equation}
 The first term measures the degree at which the magnetic moment switching is achieved and
vanishes in the case of full switching. The second term is quadratic in the driving field and is
thus proportional to the absorbed energy. The parameter $\eta$, called the \textit{control
parameter}, allows us to balance the second condition with respect to the first.

Therefore, the problem of optimal control boils down to minimizing
the cost functional (\ref{eq:CostFunctional}) along the trajectory
given by DLLE (\ref{eq:DrivenLLE}). More explicitly, this amounts to solving the following
problem

%
\begin{equation}
\left\{ \begin{array}{l}
\min\left\{ \mathcal{F}\left[\mathbf{s,h _{\mathrm{AC}}}\right]=\frac{1}{2}\left\Vert
\mathbf{s(}\tau_{f}\mathbf{)-s}^{(f)}\right\Vert ^{2}+\frac{\eta}{2}\int\limits
_{0}^{\tau_{f}}d\tau\,\mathbf{h} _{\mathrm{AC}}^{2}(\tau)\right\} \\\\ 
\mathbf{\dot{s}}=-\mathbf{s}\times\mathbf{\zeta}-\alpha\,\mathbf{s}\times\left(\mathbf{s}\times\mathbf{\zeta}\right),
\quad\tau\in\left[0,\tau_{f}\right]\\\\
\mathbf{s(}0)=\mathbf{s}^{(i)}.\end{array}\right.
\label{eq:OptimalControlProblem}
\end{equation}
%

An optimal solution of this problem is characterized by the first
order optimality condition in the form of the Pontryagin minimum principle
(PMP) \cite{pontryaginetal62}.
These conditions are more conveniently formulated with the help of a Hamilton function which may be
in the present case written in the following form
\begin{eqnarray}
\mathcal{H}\left[\mathbf{s}(\tau),\mathbf{\lambda}(\tau),\mathbf{h} _{\mathrm{AC}}(\tau)\right]&=&\frac{\eta}{2}
\mathbf{h} _{\mathrm{AC}}^{2}(\tau) \\ \nonumber
&+&\mathbf{\lambda}(\tau)\cdot\left\{
-\mathbf{s}\times\mathbf{\zeta}-\alpha\mathbf{s}\times\left(\mathbf{s}\times\mathbf{\zeta}
\right)\right\} ,\label{eq:HamiltonFunction}
\end{eqnarray}
where $\mathbf{\lambda}(\tau)$, called the adjoint state
variable {[}see below{]}, is a Lagrange parameter introduced
to implement the constraint and thereby render $\mathbf{s}(\tau)$
independent of $\mathbf{h} _{\mathrm{AC}}(\tau)$. The PMP then states that solving
the problem (\ref{eq:OptimalControlProblem}) is equivalent to solving
the following boundary problem (\textit{i.e.}, the Hamilton-Jacobi equations
with boundary conditions)
\begin{equation}
\left\{ \begin{array}{l}
\dot{\mathbf{s}}=\frac{\delta\mathcal{H}}{\delta\mathbf{\lambda}},\quad\mathbf{s}(0)=\mathbf{s}^{(i)
},\quad\tau\in\left[0,\tau_{f}\right],\\\\ 
\dot{\mathbf{\lambda}}=-\frac{\delta\mathcal{H}}{\delta\mathbf{s}},\quad\mathbf{\lambda}(\tau_{f}
)=\mathbf{s}(\tau_{f})-\mathbf{s}^{(f)},\\\\ 
\frac{\delta\mathcal{H}}{\delta\mathbf{h} _{\mathrm{AC}}}=0.\end{array}\right.\label{eq:PMP_BVP}
\end{equation}
The last condition is also equivalent to the vanishing of the gradient of the cost functional
$\mathcal{F}$ in Eq. (\ref{eq:CostFunctional}). It yields the equation
\begin{equation}
\frac{\delta\mathcal{H}}{\delta\mathbf{h} _{\mathrm{AC}}}=\eta\, \mathbf{h} _{\mathrm{AC}}+\mathbf{s}\times\mathbf{\lambda}-\alpha\,
\mathbf{s}\times\left(\mathbf{s}\times\mathbf{\lambda}\right),\label{eq: explicit 
gradient}
\end{equation}
which is used to compute the variation in the cost functional, that is
\begin{equation}
\delta\mathcal{F}=\int\limits
_{0}^{\tau_{f}}d\tau\,\frac{\delta\mathcal{H}}{\delta\mathbf{h} _{\mathrm{AC}}}\cdot\delta\mathbf{h} _{\mathrm{AC}}.
\label{eq:CostFuntlVariation}
\end{equation}

 In general, this problem is highly nonlinear and considering, on
top of that, the non-linearity of the Landau-Lifshitz equation, it
is not possible to find analytical solutions. Consequently, we
resort to numerical approaches. The advantage of this formulation
is manifold: i) the MW field $\mathbf{h} _{\mathrm{AC}}(\tau)$ is obtained
in $3D$, \textit{i.e.}, one obtains the three functions of time $h_{\mathrm{AC}}^{\alpha}(\tau),\alpha=x,y,z$;
and for any potential energy (anisotropy, DC field, etc), ii) the
final time $\tau_{f}$, and the absorbed power (second term in Eq. (\ref{eq:CostFunctional})) can
be adjusted; the latter may be achieved by tuning the control parameter
$\eta$, iii) one can generalize this treatment to many-spin problems \cite{kacgar05springer} and
also include thermal effects.

One of the most efficient techniques for (numerically) solving
such a minimization problem is the conjugate-gradient method. However,
the drawback of this method is that it is a local-convergence method,
which means that the solution it renders is strongly dependent on
the initial guess. We overcome this inconvenience
by supplementing the method by a global search using the Metropolis algorithm
with random increments and then proceed by the technique of simulated
annealing.

For numerical calculations, we have discretized the boundary-value
problem (\ref{eq:PMP_BVP}) by subdividing the time interval $\left[\tau_{i}=0,\tau_{f}\right]$
into $N$ time slices
\begin{align*}
\tau_{n} & =\tau_{i}+n\times\Delta\tau,\quad n=0,\ldots,N-1,\quad\tau_{f}=\tau_{N-1},
\end{align*}
where
\[
\Delta\tau=\frac{\tau_{f}-\tau_{0}}{N-1}.
\]

Then, using the notation $\mathbf{v}_{n}=\mathbf{v}(\tau_{n})$ for
a vector $\mathbf{v}$, Eqs. (\ref{eq:DrivenLLE}, \ref{eq:CostFunctional},
\ref{eq:CostFuntlVariation}) and the equation for $\mathbf{\lambda}$,
become

\begin{widetext} 
\begin{subequations} \label{eq:DiscreteOCP}
\begin{eqnarray}
 &  &
\mathbf{s}_{n+1}=\mathbf{s}_{n}+\Delta\tau\times\left[-\mathbf{s}_{n}\times\mathbf{\zeta}_{n}-\alpha
\,\mathbf{s}_{n}\times\left(\mathbf{s}_{n}\times\mathbf{\zeta}_{n}\right)\right],\quad\mathbf{s}
(\tau_{i})=\mathbf{s}^{(i)},\\ 
 &  & \mathcal{F}=\frac{1}{2}\left\Vert \mathbf{s}_{N-1}\mathbf{-s}^{(f)}\right\Vert
^{2}+\frac{\eta\Delta\tau}{2}\sum\limits _{n=0}^{N-1}\mathbf{h} _{\mathrm{AC},n}^{2},\\ 
 &  &
\mathbf{\lambda}_{n-1}=\mathbf{\lambda}_{n}-\Delta\tau\times\Lambda_{n},\quad\mathbf{\lambda}_{f}
=\mathbf{s}_{N-1}-\mathbf{s}^{(f)},\\ 
 &  &
V_{n}=\frac{\delta\mathcal{F}}{\delta\mathbf{h} _{\mathrm{AC},n}}=\Delta\tau\times\left[\eta\,\mathbf{h} _{\mathrm{AC},n}
+\mathbf{s}_{n}\times\mathbf{\lambda}_{n}-\alpha\,\mathbf{s}_{n}\times\left(\mathbf{s}_{n}
\times\mathbf{\lambda}_{n}\right)\right].
\end{eqnarray} 
\end{subequations}
\end{widetext}

The explicit expression for $\Lambda_{n}$ in (\ref{eq:DiscreteOCP}c) depends on the energy
potential [see below for the case of uniaxial anisotropy].

We may summarize the numerical procedure as follows. i) for a given
initial guess of the control field $\mathbf{h} _{\mathrm{AC}}(t)$, we first solve
the state equation (\ref{eq:DiscreteOCP}a) forward in time using the initial condition, and then evaluate the cost functional
(\ref{eq:DiscreteOCP}b), ii) the solution obtained for $\mathbf{s}$
is then used for the backward (since the condition now is at $t_{f}$)
integration of the equation (\ref{eq:DiscreteOCP}c) for $\mathbf{\lambda}$,
iii) with the trajectories of $\mathbf{s}$ and $\mathbf{\lambda}$ thus obtained
we compute the gradient (\ref{eq:DiscreteOCP}d). The numerical subroutines
are standard and can be found in Ref.~\onlinecite{pressetal02}.
We emphasize that obtaining the control field amounts to solving for
$3\times N$ variables.

\paragraph{Uniaxial anisotropy}

In the case of uniaxial anisotropy with oblique static field
the energy of the nanomagnet reads (in units of the anisotropy energy $2KV$)
\begin{equation}
\mathcal{E}=-h _{\mathrm{DC}}\,\left(\mathbf{e}_{h}\cdot\mathbf{s}\right)-\frac{1}{2}(\mathbf{s}\cdot\mathbf{n})^
{2},\label{NanomagnetEnergy}
\end{equation}
with $K$ and $\mathbf{n}$ being the anisotropy constant and easy
axis, $V$ the nanomagnet volume and $h _{\mathrm{DC}}\equiv H _{\mathrm{DC}}/H_{a}$. The effective
field explicitly reads [see Eq. \ref{eq:EffRedField}]
\begin{equation}
\mathbf{h}_{\mathrm{eff}}=h _{\mathrm{DC}}\,\mathbf{e}_{h}+(\mathbf{s}\cdot\mathbf{n})\,\mathbf{n}.
\label{ReducedEffectiveField}
\end{equation}
From the second equation in (\ref{eq:PMP_BVP}) we obtain the explicit equation for
$\mathbf{\mathbf{\lambda}}$ 
\begin{eqnarray}
\dot{\mathbf{\lambda}} & = &
\mathbf{\zeta}\times\mathbf{\lambda}+\alpha\left[\mathbf{\zeta}\times\left(\mathbf{\lambda}
\times\mathbf{s}\right)+\mathbf{\lambda}\times\left(\mathbf{\zeta}\times\mathbf{s}\right)\right]
\nonumber \\ 
 & + &
\left[\mathbf{\lambda\cdot}\left(\mathbf{s}\times\mathbf{n}+\alpha\,\mathbf{s}\times\left(\mathbf
{s}\times\mathbf{n}\right)\right)\right]\mathbf{n}\label{eq:LambdaEqt}
\end{eqnarray}
and in Eq. (\ref{eq:DiscreteOCP}c) we now have
\begin{align*}
\Lambda_{n} &
=\mathbf{\zeta}_{n}\times\mathbf{\lambda}_{n}+\alpha\left[\mathbf{\zeta}_{n}\times\left(\mathbf{
\lambda}_{n}\times\mathbf{s}_{n}\right)+\mathbf{\lambda}_{n}\times\left(\mathbf{\zeta}_{n}
\times\mathbf{s}_{n}\right)\right]\\ 
 &
+\left[\mathbf{\lambda}_{n}+\alpha\left(\mathbf{\lambda}_{n}\times\mathbf{s}_{n}\right)\right]
\cdot\left(\mathbf{s}_{n}\times\mathbf{n}\right)\mathbf{n}.
\end{align*} 

\section{Results}

In the present work, we have considered the case of a nanomagnet with
uniaxial anisotropy and oblique static field. Unless otherwise stated,
the latter is applied in the $yz$ plane making an angle of $170\text{\textdegree}$
with respect to the easy axis ($z$ axis). Its reduced magnitude is $h _{\mathrm{DC}}=0.5$, corresponding to a 
field magnitude $H _{\mathrm{DC}} \simeq 150$ mT. The initial position and target states 
$\mathbf{s}^{(i)}$ and $\mathbf{s}^{(f)}$, which correspond respectively to the metastable equilibrium state 
and stable equilibrium state, are computed numerically. 
The observation time is $\tau_{f}=600$ (\textit{i.e.}, $t _f \simeq 11.4 $ ns). The damping parameter is $\alpha=0.05$
and the control parameter $\eta$ has been set to $0.01$. 
The static field, damping parameter and observation time have been varied and 
their effects studied [see later on].
For simplicity, we have taken a linearly polarized MW field, \textit{i.e.},
$\mathbf{h} _{\mathrm{AC}}(t) = h _{\mathrm{AC}}(t)\mathbf{e}_{x}$. This choice also suits
the experimental setup\cite{microSQUID}.

\begin{figure}[floatfix]
 \includegraphics[width=0.95\columnwidth]{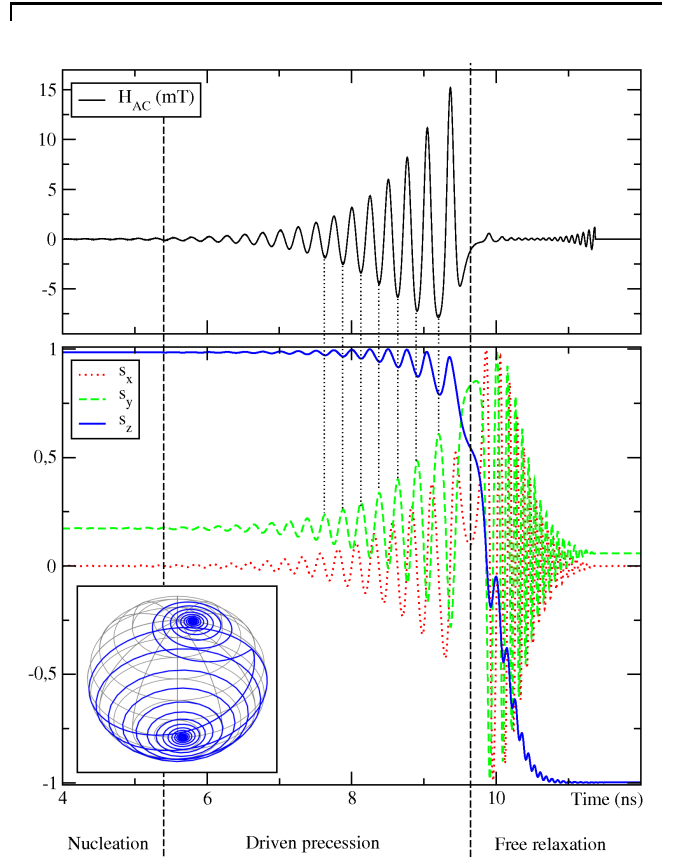} 
\caption{\label{fig:SpinTrajRFField}Optimized MW field (upper panel) and the corresponding spin
trajectories (lower panel). The inset is a $3D$ plot of the spin trajectory on the unit sphere.}
\end{figure}

\begin{figure}[floatfix]
 \includegraphics[width=0.95\columnwidth]{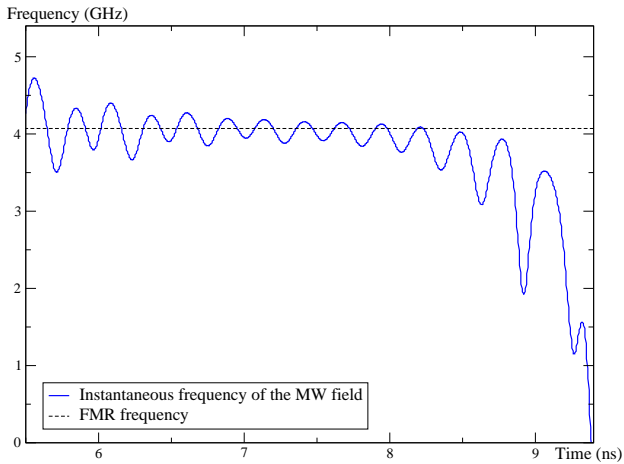}
\caption{\label{fig:TimeDepFreq} Instantaneous frequency of the optimized and filtered MW field of 
Fig. \ref{fig:SpinTrajRFField}. The other parameters are the same as in Fig. \ref{fig:SpinTrajRFField}.}
\end{figure}

In Fig. \ref{fig:SpinTrajRFField} we have plotted the optimized MW field magnitude 
$ H _{\mathrm{AC}}(t) \equiv H_a\,h _{\mathrm{AC}}\left(t\right)$ where $t$
is the time in seconds, together with the components of the magnetic moment, \textit{i.e.},
$s_{\alpha}\left(t\right),\alpha=x,y,z$.
First, we note that the amplitude of the MW field is rather small as it does not exceed $15$
mT, which is $10$ times smaller than the static field. Moreover, the summed magnitudes of the DC and MW 
field are smaller than the SW switching field for the chosen DC field direction (about $200$ mT). 
This shows that, in the presence of a MW field, magnetic switching is achieved at a smaller DC field. 
Second, the striking feature is that the MW field is modulated both in amplitude and frequency. 
Its frequency is a slowly varying function of time in the stage that precedes switching,
as can be seen in Fig. \ref{fig:TimeDepFreq}.
Third, as is hinted to by the dashed vertical lines, the extrema in the MW field and the spin
components $s_y(t)$ and $s_z(t)$ match at all times before switching. This simply implies that the magnetic
moment is phase-locked to the MW field.
All these features agree with the predictions of the classical auto-resonance or the ladder-climbing
quantum theory, as summarized in the introduction.

The instantaneous frequency has been obtained after passing the optimized MW field through the 
Butterworth filter and then applying the Hilbert transformation
\cite{girvak02mst,goshoe04sp}. 
As can be seen, for short times the instantaneous frequency oscillates around the approximate value
$f_{0}\approx4.1\,\mbox{GHz}$. This initial frequency is simply the FMR frequency given by
$$
f_\mathrm{FMR} = \frac{\gamma H_{a}}{2 \pi }
\sqrt{
h_{\mathrm{eff}, \parallel}^{(i)}
\left(
h_{\mathrm{eff}, \parallel}^{(i)}
+
k\left[\left(\mathbf{s}^{(i)}.\mathbf{n}\right)^{2} - 1\right]
\right)
}
$$
where
$
 h_{\mathrm{eff}, \parallel}^{(i)} \equiv
 \mathbf{h _{\mathrm{eff}i}}^{(i)}\cdot\mathbf{s}^{(i)} =
 h _{\mathrm{DC}} \left(\mathbf{e}_{h}\cdot\mathbf{s}^{(i)}\right) + (\mathbf{s}^{(i)}\cdot\mathbf{n})^2
$
is the effective field (\ref{ReducedEffectiveField}) evaluated at and then projected onto the
initial position $\mathbf{s}^{(i)}$.
As the magnetic moment approaches the saddle point the frequency decreases rapidly
and eventually vanishes when the magnetic moment crosses the saddle point into the more stable
energy minimum.

In Fig. 1 it is seen that the time span comprises three stages (for the set of physical
parameters considered): 1) \emph{Nucleation stage} (up to $5.4\,\mathrm{ns}$).
The MW field remains almost zero and the magnetic moment remains in
the metastable state. 2) \emph{Driven precession} (from $5.4\,\mathrm{ns}$
to $9.7\,\mathrm{ns}$). Here the MW field and the magnetic moment are
synchronized. At each procession cycle, the MW field hooks up the magnetic
moment and pushes it upwards in the energy potential towards the saddle point.
This is the phase-locking process mentioned in the introduction and observed above. This is
indeed possible because the frequency chirp rate is small as can be seen in Fig.
\ref{fig:TimeDepFreq} for $5.4\,\mathrm{ns}\leq t\leq 9.7\,\mathrm{ns}$.
The MW field thus compensates for the effect of damping that tends to pull
the magnetic moment back towards its initial position. At around
$9.7\,\mathrm{ns}$, the magnetic moment crosses the saddle point. 3)
\emph{Free relaxation}: from $9.7\,\mathrm{ns}$ onward, the magnitude
of the MW field dwindles and the synchronization with the magnetic
moment is lost.
We note that at the saddle point the precession reverses from being counter-clockwise to clockwise
as the magnetic moment switches to the lower half sphere.

Numerical tests show that the MW field can be replaced by zero during
the nucleation and free relaxation stages without noticeably affecting
the trajectory of the magnetic moment. This implies that the most
relevant part of the signal is that during the driven precession;
the role of the MW field is thus to drive the magnetic moment
towards the saddle point. Next, the damping takes up to lead it to
the more stable energy minimum.
During the driven precession the frequency of the MW field and the
precession frequency of the magnetic moment are similar. Consequently, the magnetic moment
switching can be viewed as a resonant process: the pumping by the
MW field is efficient when its frequency matches the frequency of
the magnetization (phase-locking).
\begin{figure}[floatfix]
 \includegraphics[width=0.95\columnwidth]{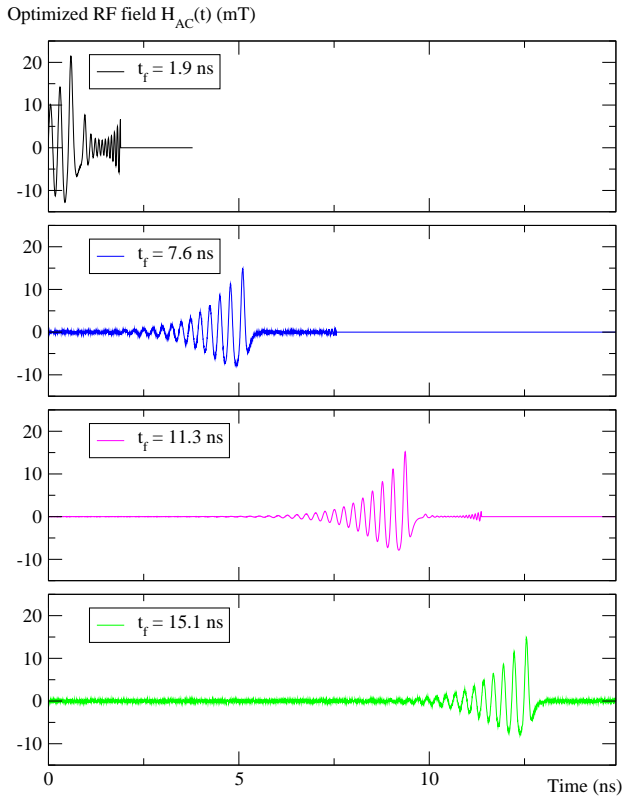}
\caption{\label{fig:EffectObsTime}Optimized MW field obtained with four different total times
$t_f$.}
\end{figure}

The same calculation has been carried out with the same sampling time but different values for the
total observation time $t_f$. The results are shown in Fig. \ref{fig:EffectObsTime}. 
If the total time is larger than an effective time of $6\,\mathrm{ns}$, similar values are obtained 
for the cost functional and the
curves $h _{\mathrm{AC}}(t)$ can be matched after a time shift. As was discussed earlier, this effective time
corresponds to the sum of the time of driven precession and that of free relaxation. 
This result implies that the nucleation stage can be suppressed without affecting the final
optimized MW field. However, if the total time is too short, the final value found for the
cost functional is higher (\textit{i.e.} not fully minimized). Indeed, we see in Fig. \ref{fig:EffectObsTime} 
(uppermost panel) that the stage of driven precession is shortened and the shape of the control 
field changes so as to achieve a faster switching and thus comply with the switching-time constraint 
[first term in Eq. (\ref{eq:CostFunctional})].
\begin{figure}[floatfix]
\includegraphics[width=0.95\columnwidth]{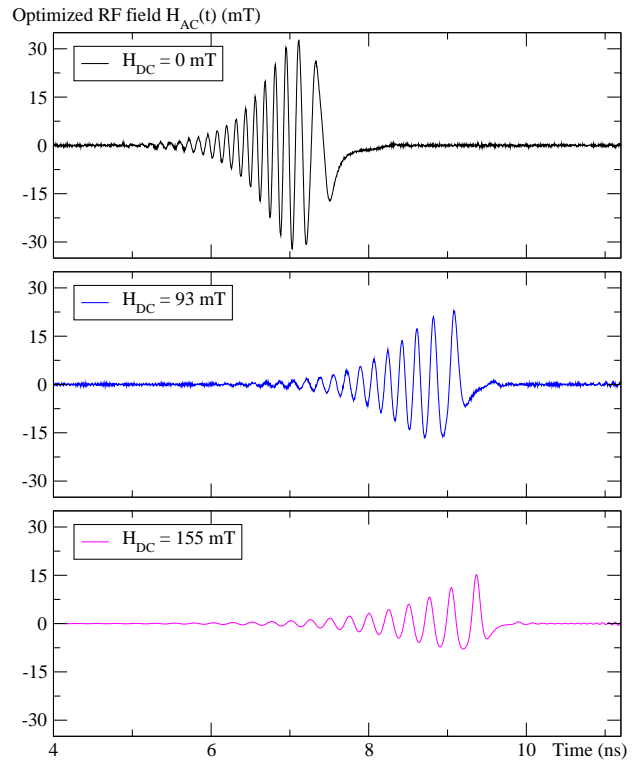}
\caption{\label{fig:EffectDCFieldAmp}Optimized MW field obtained for different
magnitudes of the static field $\mathbf{h} _{\mathrm{DC}}$, in the same direction
making an angle $\theta=170\text{\textdegree}$ with respect to the
anisotropy easy axis.}
\end{figure}

The effect of varying the amplitude of the static field on the MW field is
shown in Fig. \ref{fig:EffectDCFieldAmp}. We see that the shape of
the MW field envelop remains the same, apart from the fact that the
smaller the static field, the more symmetrical is the MW field.
This shows that for a higher field $h _{\mathrm{DC}}$, the energy potential
is less symmetrical.
Moreover, as $h _{\mathrm{DC}}$ is increased the energy barrier is lowered and the MW field required to achieve 
switching is smaller. Again, the initial frequency of the oscillations matches the FMR frequency of
the system; when $h _{\mathrm{DC}}$ increases, the latter decreases.
The cost functional was found to be proportional to the energy barrier
between the saddle point and the metastable minimum. Hence, when the
energy barrier is higher, more energy has to be injected in order
to overcome it. The same study has been carried out upon varying the
direction of the static field.
\begin{figure}[floatfix]
 \includegraphics[width=0.95\columnwidth]{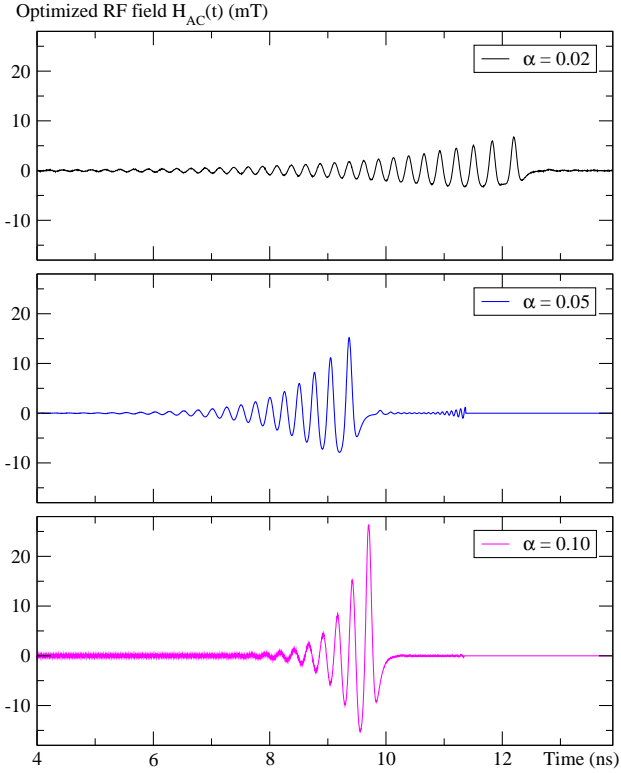}
\caption{\label{fig:EffectDamping}Optimized MW field obtained for different
values of the damping parameter $\alpha$.}
\end{figure}
\begin{figure}[floatfix]
 \includegraphics[width=0.95\columnwidth]{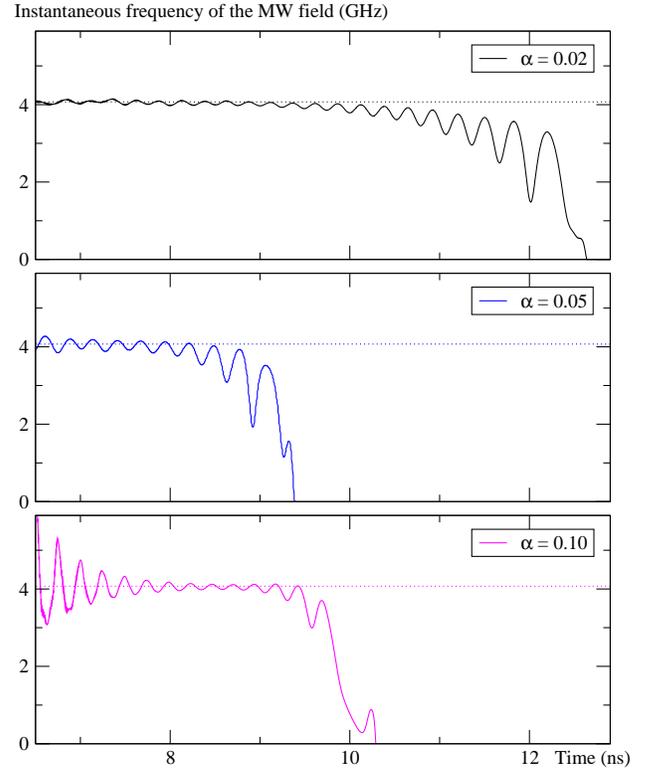}
\caption{\label{fig:EffectDampingFrequency}Instantaneous frequency of the MW field optimized for several
values of the damping parameter $\alpha$. Dotted line: FMR frequency.}
\end{figure}

We have also investigated the effect of varying the damping parameter $\alpha$
on the MW field. The results are summarized in Fig. \ref{fig:EffectDamping}.
We see that the intensity of the field increases with $\alpha$, which is compatible with what was 
suggested earlier, namely that the role of the MW field is to compensate for the damping effect. 
This effect is similar to what happens with a rubber band: the more you stretch it the harder it becomes to do so. 
Moreover, the effective duration of the MW field, which mainly corresponds to the driven
precession period, decreases when $\alpha$ increases. We note that, on the contrary, the initial
frequency of the oscillations is independent of $\alpha$ [see Fig. \ref{fig:EffectDampingFrequency}].
This result can be understood qualitatively if we suppose that, at any time, the MW field exactly 
compensates for the effect of damping. The spin dynamics is then governed by the undamped LLE  
and the magnetic moment precesses with its proper frequency, which is independent of the damping parameter. 
At short times, since the precession angle is small, this precession frequency is equal to the FMR frequency. 

As discussed in the introduction, one of the objectives of investigating
the magnetization switching assisted by MWs is to achieve an
optimal switching with smaller DC magnetic fields than it would be
necessary without MWs.
This means that applying the DC field in a given direction and varying its magnitude one determines
the switching field (or the field at the limit of metastability) at which the magnetization is
reversed.
This is the SW astroid. Due to
the energy brought into the system by MWs, the field required for switching
is smaller. This has been nicely demonstrated using the $\mu$-SQUID technique on a 20-nm cobalt
particle \cite{microSQUID}.
The  most striking feature of the SW astroid obtained by
these measurement is its jaggedness. In other words, the reduction
of the switching field is not uniform and presents a kind of {}``fractal''
character.
The global features depend on several physical parameters,
such as the MW field pulse duration, its rising time, its frequency,
the DC field amplitude, and the damping parameter. In the present
work, and in the particular case considered here, namely that of uniaxial
anisotropy, we first wanted to check whether this reduction of the
switching field is recovered by our optimal-control method.
Furthermore, we address the question as to whether the SW astroid may be used as a fingerprint of a 
given nanocluster. More precisely, the question is whether a given SW astroid can provide us with
specific information about the corresponding cluster, like its energy potential and the physical
parameters such as damping.

\begin{figure}[floatfix]
 \includegraphics[width=0.95\columnwidth]{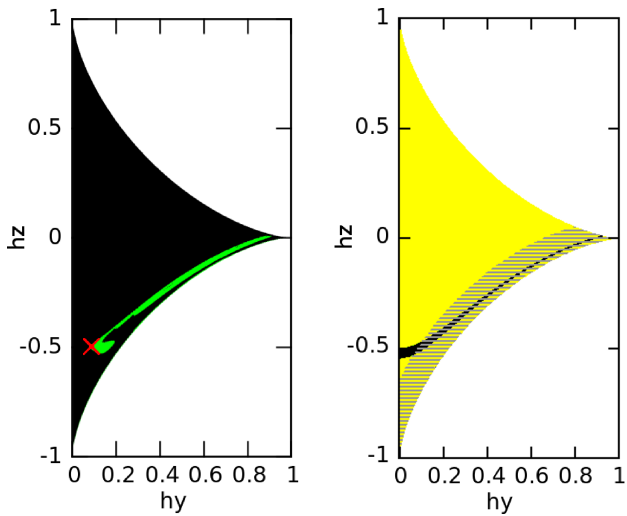} 
\caption{\label{fig:SwitchingCurves}(a) Switching curve computed in the presence
of the MW field $\mathbf{h} _{\mathrm{AC}}^{0}(t)$. The red cross indicates the amplitude and direction 
of the DC field for which the MW field was optimized. The area in green
is where switching has been achieved, the black area is where there
is no switching, and in the white area the static field is higher
than the switching field (\textit{i.e.} beyond the metastability region). 
(b) In the black area the FMR frequency is the same as for the reference DC field 
$\mathbf{h} _{\mathrm{DC}}^{0}$. In the hatched area the energy barrier between the metastable 
minimum and the saddle point (computed numerically) is smaller than for the reference 
DC field $\mathbf{h} _{\mathrm{DC}}^{0}$.
 }
\end{figure}

Accordingly, we check whether an MW field $\mathbf{h} _{\mathrm{AC}}^{0}(t)$, which is optimized in the presence 
of a reference applied DC field $\mathbf{h} _{\mathrm{DC}}^{0}(t)$ with given direction and magnitude, 
\textit{e.g.} $h _{\mathrm{DC}}=0.5$ and an angle of $170\text{\textdegree}$ with respect to the easy axis, 
can still induce magnetization switching in the presence of another 
DC field, with different direction and/or magnitude.
To answer this question, the MW field $\mathbf{h} _{\mathrm{AC}}^{0}(t)$, was used in the driven
LLE (\ref{eq:DrivenLLE}) and the calculation of the switching field was performed for several
intensities and directions of the static field $\mathbf{h}$ leading to the switching curves in
presence of $\mathbf{h} _{\mathrm{AC}}^{0}(t)$ as shown in Fig. \ref{fig:SwitchingCurves}.
As can be seen, the magnetization switching occurs only inside the golf-club-shaped green area
[see Fig. \ref{fig:SwitchingCurves} (left)]. In the black area, the pumping by the MW field is 
inefficient and switching does not occur. This curve is in agreement with the experimental data 
of Ref.~\onlinecite{microSQUID}.

The shape of the green pattern can be explained based on qualitative arguments about the frequency and magnitude 
of the MW field. As has been seen previously, in order to achieve the switching, the MW field must fulfill 
the following conditions: i) it must be synchronized with the proper precession frequency of the magnetization; 
so at short times its frequency must match the FMR frequency of the system, and ii) the injected energy,
must be sufficient to overcome the energy barrier between the metastable minimum and the saddle point. 

For any magnitude or direction of the field $\mathbf{h} _{\mathrm{DC}}$, both the FMR frequency and the energy barrier 
can be computed numerically [see Fig. \ref{fig:SwitchingCurves} (right)]. In the black area the value of the 
FMR frequency is the same as for $\mathbf{h} _{\mathrm{DC}}^{0}$. In the hatched area 
the energy barrier is lower than for the $\mathbf{h} _{\mathrm{DC}}^{0}$. Outside the black zone, the MW field is not 
synchronized with the precession frequency of the system: the switching can not occur. Outside the hatched area 
the injected energy is not sufficient to overcome the energy barrier. Consequently, the switching is only 
achieved in the intersection between both areas. Indeed, comparing with Fig. \ref{fig:SwitchingCurves} (left), 
this intersection matches more or less the green zone, where the switching occcurs. 

\begin{figure*}[floatfix]
\includegraphics[width=5cm,height=5cm,keepaspectratio]{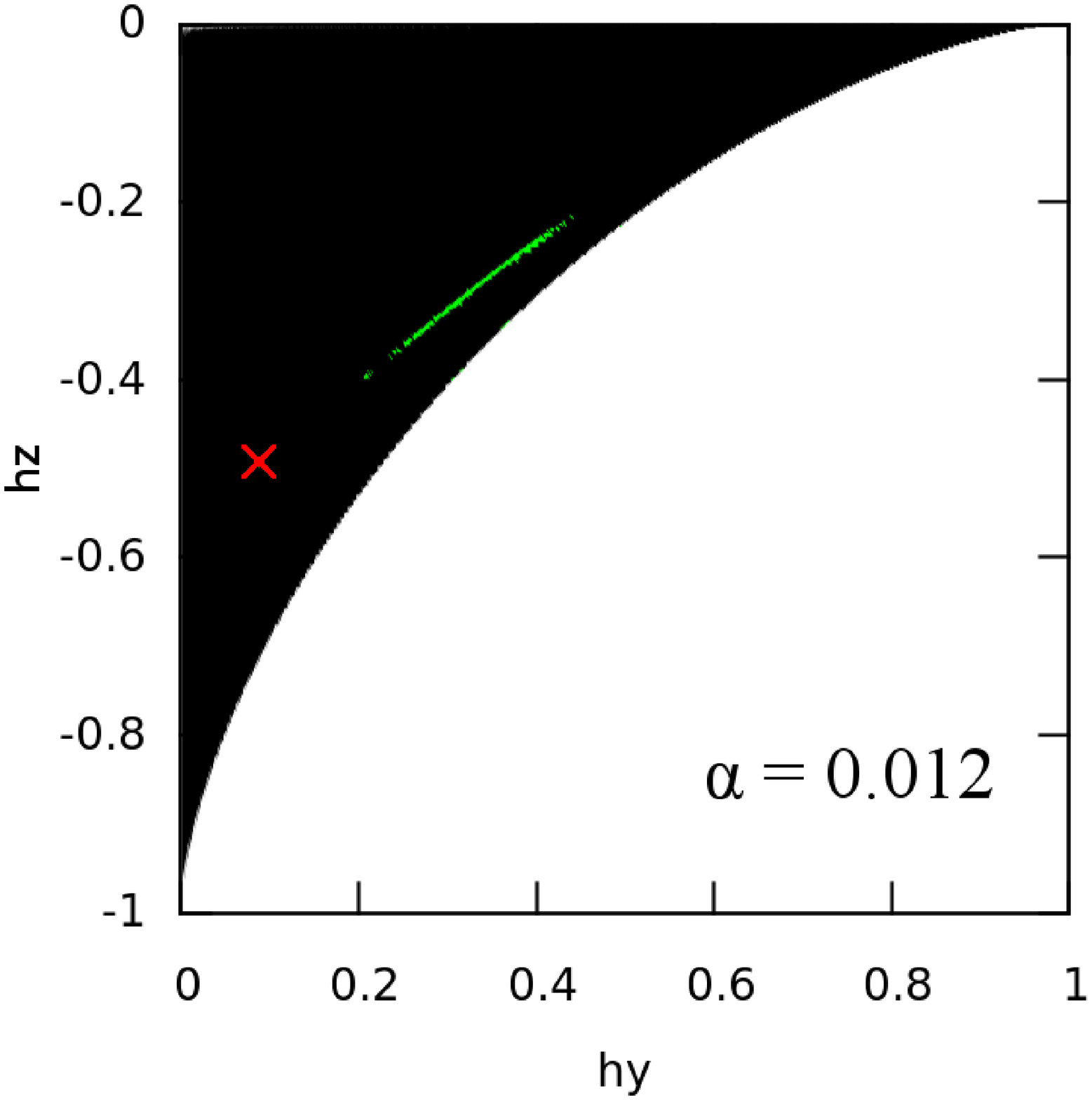}
\includegraphics[width=5cm,height=5cm,keepaspectratio  ]{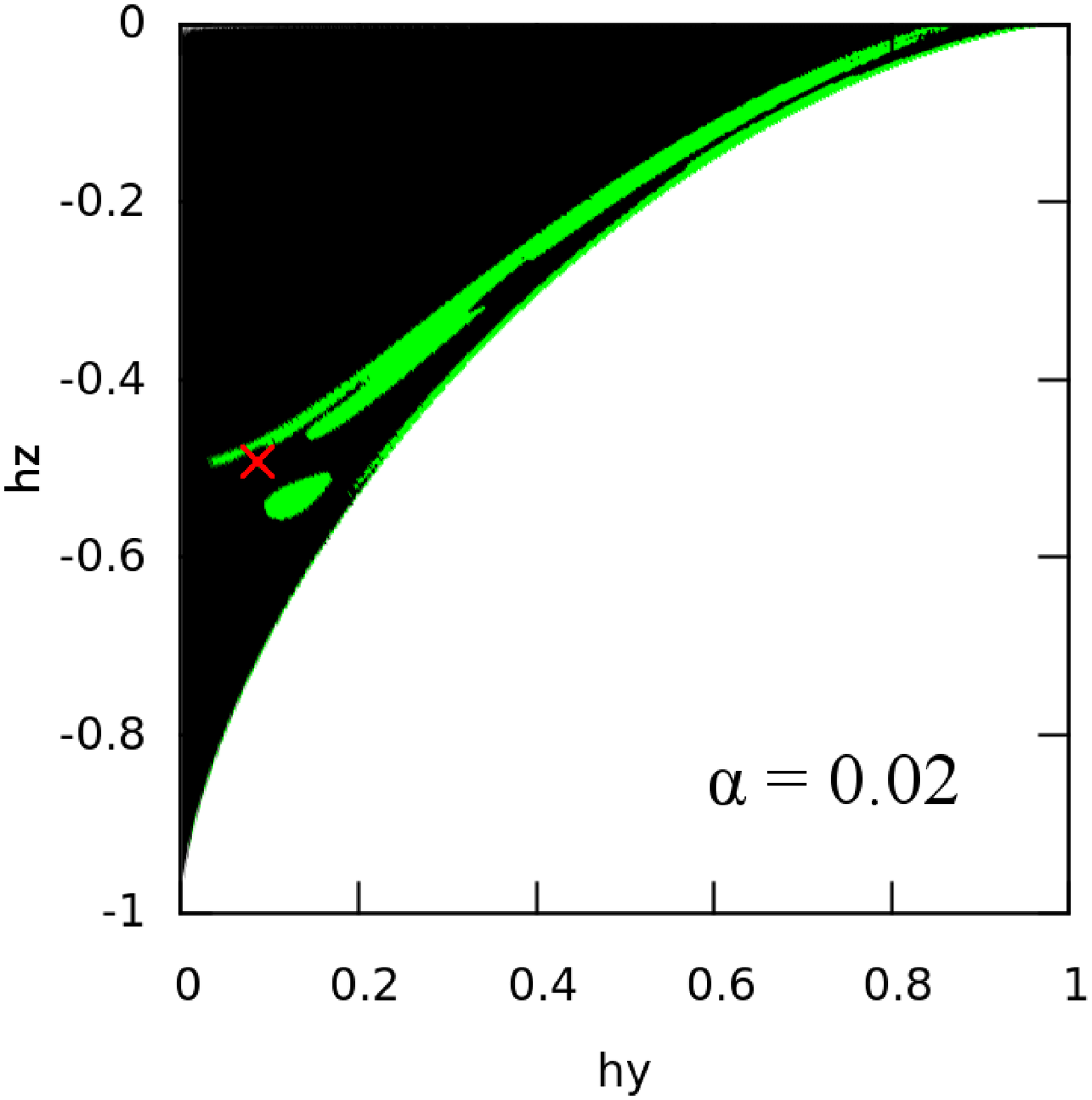}
\includegraphics[width=5cm,height=5cm]{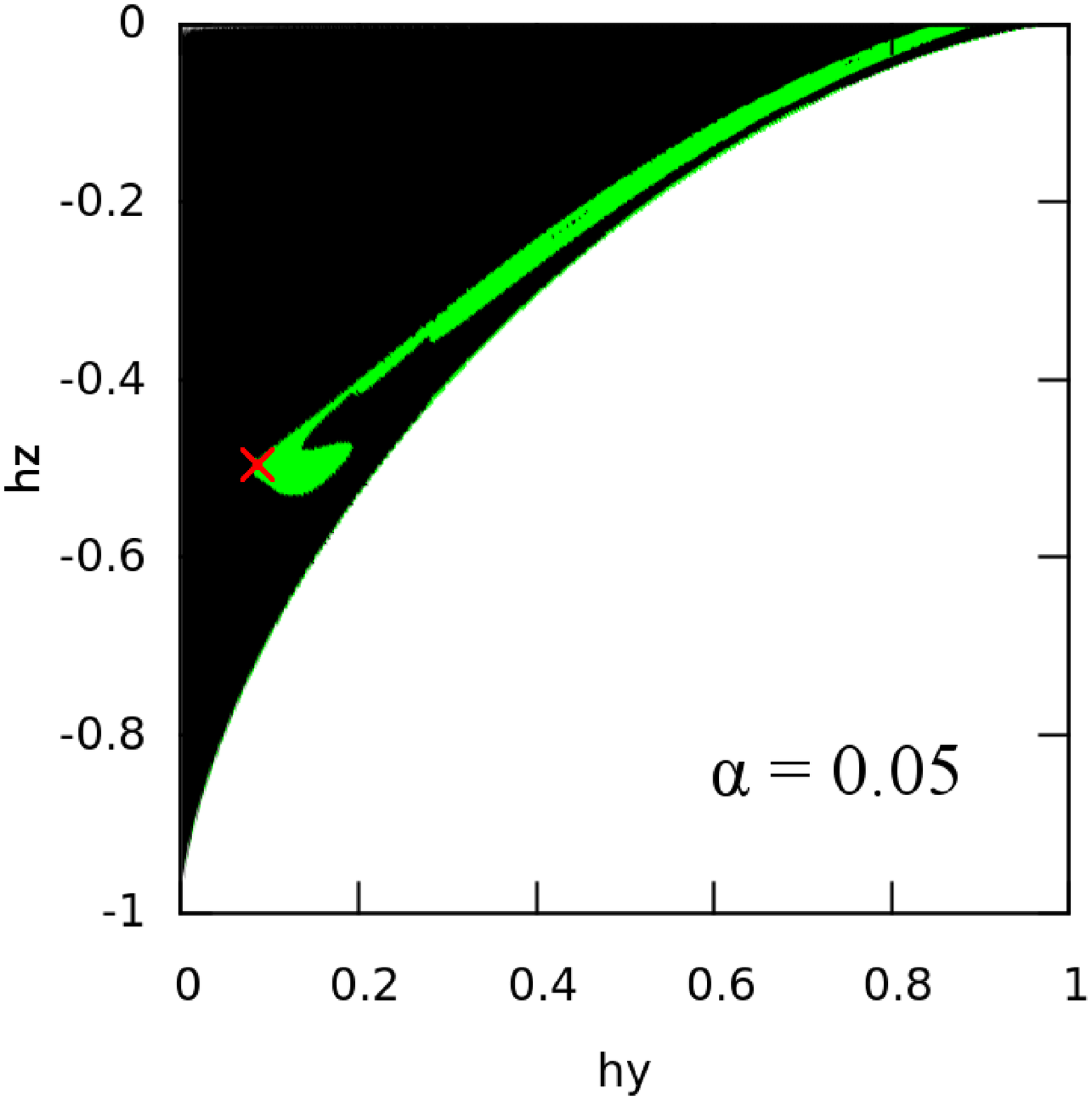}
\includegraphics[width=5cm,height=5cm,keepaspectratio]{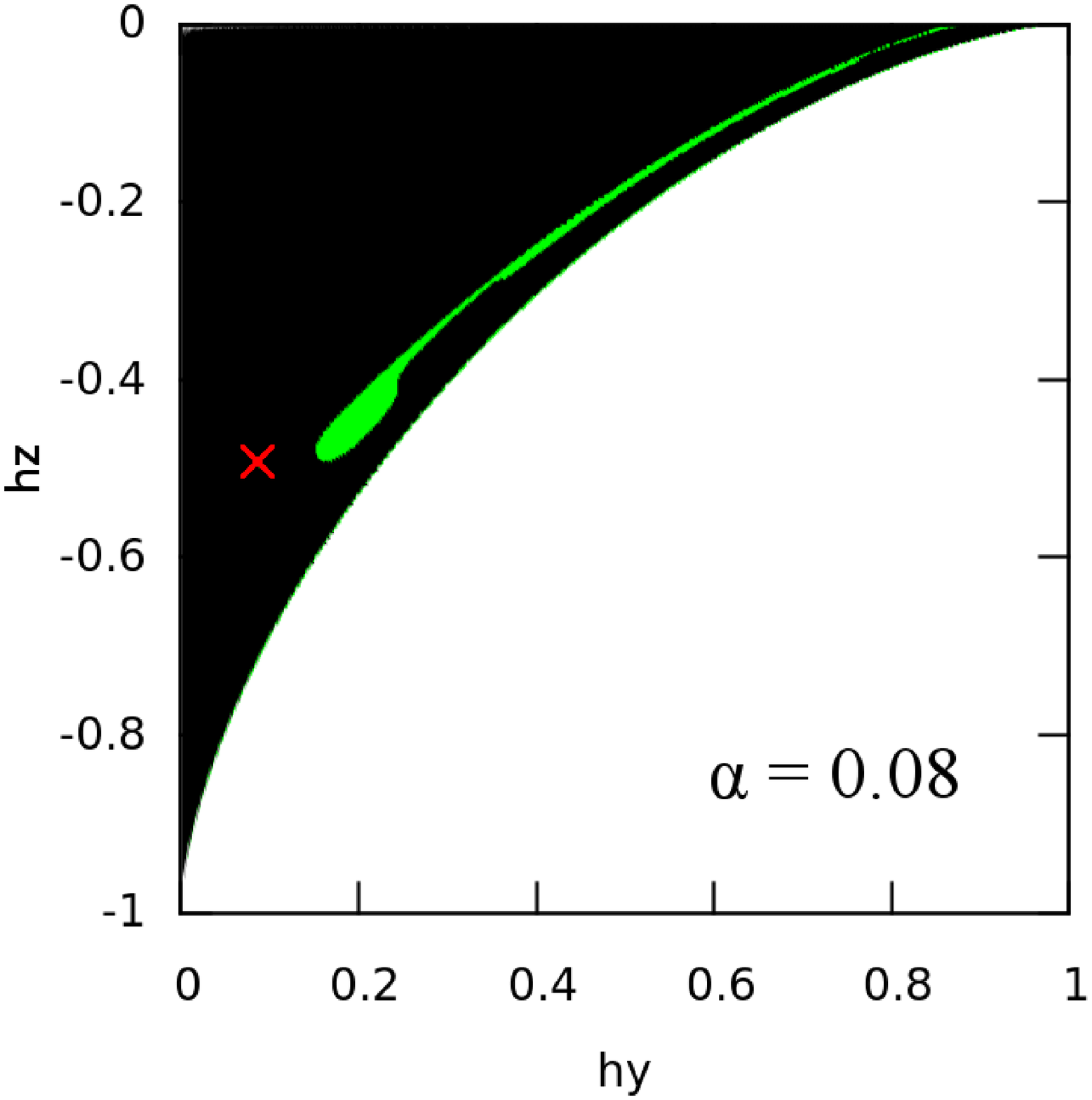}
\includegraphics[width=5cm,height=5cm,keepaspectratio]{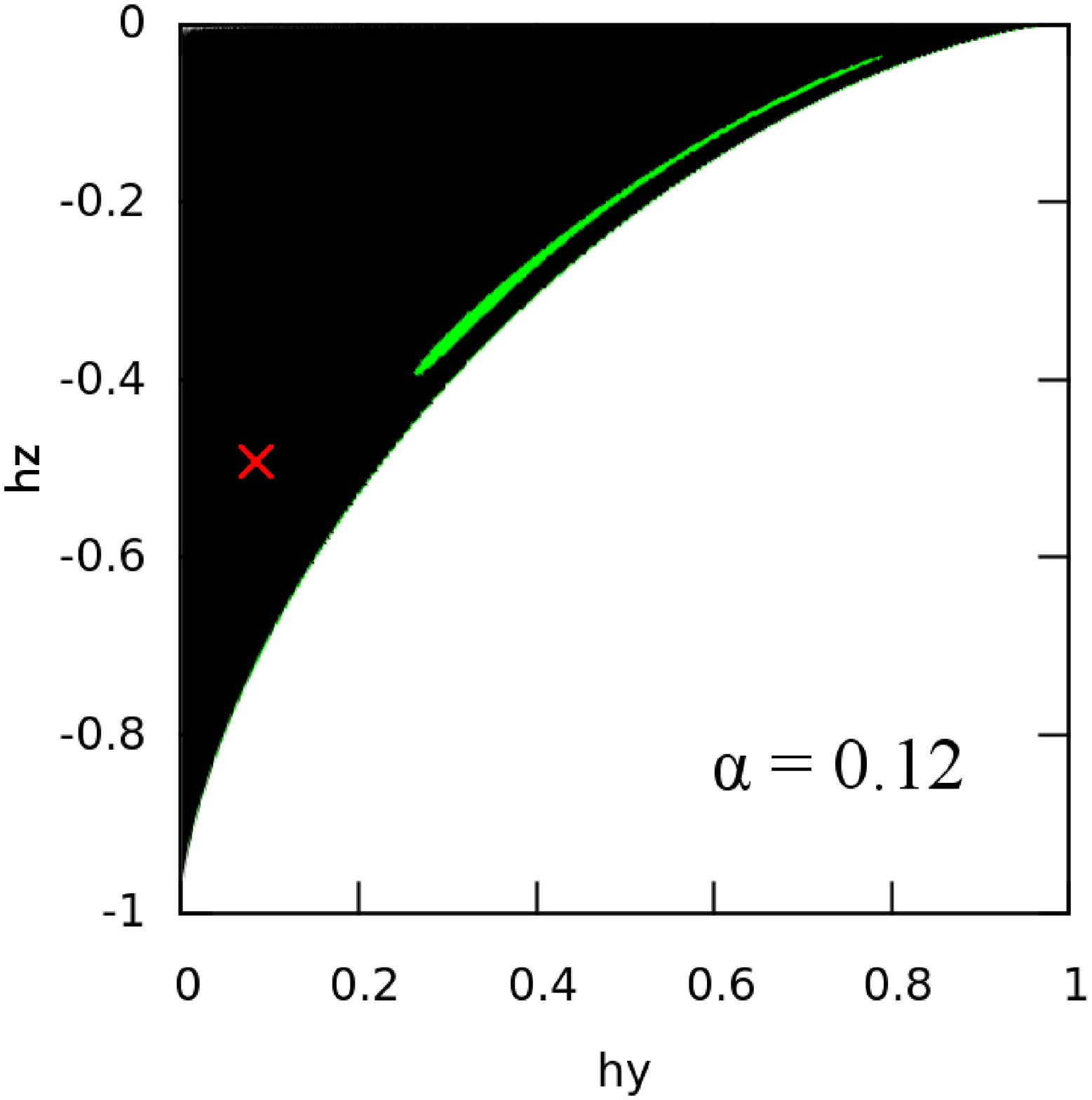}
\includegraphics[width=5cm,height=5cm,keepaspectratio]{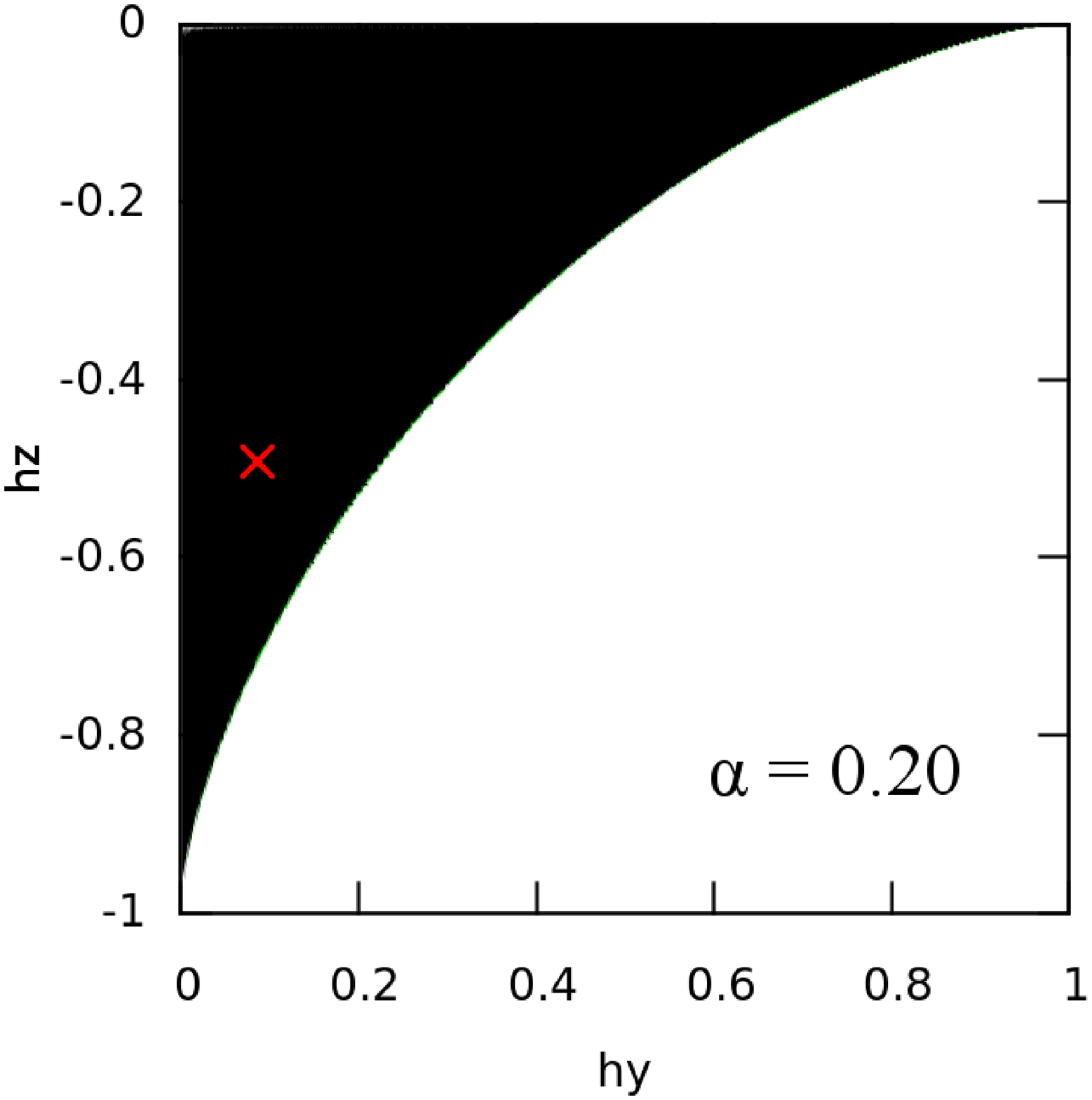}
\caption{\label{fig:SWADamp}Switching curves obtained for different values
of the damping parameter $\alpha$, with the MW 
field optimized for $h _{\mathrm{DC}}=0.5$ and an angle of $170\text{\textdegree}$ with respect to the
easy axis, and $\alpha^{0}=0.05$. The color code is similar to that in Fig. \ref{fig:SwitchingCurves}.}
\end{figure*}

Next, we optimize the MW field $\mathbf{h} _{\mathrm{AC}}^{0}(t)$ for the 
reference DC field $\mathbf{h} _{\mathrm{DC}}^{0}(t)$ with magnitude
$h _{\mathrm{DC}}=0.5$ and angle of $170\text{\textdegree}$ with respect to the easy axis, 
and damping $\alpha^{0}=0.05$; then we
compute the SW astroid for other values of $\alpha$, in the presence of the same DC and MW fields.
The results are shown in Fig. \ref{fig:SWADamp}.

We see that the shape of the switching area strongly depends on the damping parameter $\alpha$. The
largest green area is found for $\alpha=\alpha^{0}$. Then, as $\alpha$ increases the green area
shrinks and vanishes for $\alpha>0.12$. Indeed, for high values of $\alpha$, the MW field is not
strong enough to compensate for the effect of damping.
The same phenomenon is observed for small values of $\alpha$, in which case the MW field
``overcompensates'' for the effect of damping and thereby the energy can not be pumped into the
system in an efficient manner.

\section{Conclusions and outlook}

We have developed a general and efficient method for determining the characteristics (pulse
shape, duration, intensity, and frequency) of the MW field that triggers the switching of a
nanomagnet in an oblique static magnetic field. We have applied the method to the case of uniaxial
anisotropy and investigated the effect of the DC field and damping on
the optimized MW field. We have shown that our method does recover the switching field curves as
observed on cobalt nanoclusters.
It remains though to investigate the origin of the ``fractal'' character observed in the measured
switching curves.

We have shown that the MW field that triggers the magnetization switching, while minimizing the
absorbed energy, can be efficiently calculated using the optimal control theory. According to our
results, the optimal MW field is modulated both in frequency and in magnitude.
The role of this MW field is to drive the magnetization towards the saddle point, then damping
leads the magnetic moment to the stable equilibrium position. For the pumping to be efficient, the MW
field frequency must match the proper precession frequency of the magnetization, which depends on
the magnitude and the direction of the static field.
Moreover,  the intensity depends on the damping parameter. This result could be used to probe the
damping parameter in experimental nanoparticles.

The present method is quite versatile and can be extended to other anisotropies. 
It could also be used to study the dynamics of nanoclusters in the many-spin approach 
\cite{kacgar05springer}.
In this case one will probably have to deal with a nonuniform MW field,
especially if surface anisotropy is taken into account \cite{kacdim-garkac}.
One may then study switching via internal spin wave excitations
and the effect of the MW field on the corresponding relaxation rate
\cite{garkacreySWI}.
Thermal effects can also be accounted for by adding a Langevin field
on top of the DC and MW fields. In this case, it will be interesting
to investigate the interplay between the MW field and the Langevin
field and to figure out when these two fields play concomitant roles.
\begin{acknowledgments}
We are grateful to our collaborators E. Bonet, R. Picquerel, C. Thirion, W. Wernsdorfer
 (Institut N\'eel, Grenoble) and V. Dupuis (LPMCN,Lyon) 
for instructive discussion of their experiments on isolated nanoclusters. This work has been
funded by the collaborative program PNANO ANR-08-P147-36 of the French Ministry.
\end{acknowledgments}
%

%
\end{document}